\documentstyle[aps,prb,psfig,twocolumn]{revtex}  
\begin{document}
\draft

\title{Densification effects on the Boson peak in vitreous silica: \\
a molecular-dynamics study}

\author{P. Jund and R. Jullien}

\address{Laboratoire des Verres, CNRS-Universit\'e Montpellier 2,\\
Pl. E. Bataillon CC069, 34095 Montpellier, France}

\maketitle


\begin{abstract}
We perform classical molecular-dynamics simulation to study the
effect of densification on the vibrational spectrum of a model silica
glass. We concentrate this study on the so-called Boson peak and compare
our results, obtained from a direct diagonalization of the dynamical matrix, 
with experimental Raman data. We show that, upon densification, 
the position of the Boson peak shifts towards higher frequencies 
while its magnitude decreases which is in agreement with
a recent experimental study.  
\end{abstract}

\pacs{PACS numbers: 61.43.Fs, 63.50.+x, 02.70.Ns}


\narrowtext

\section{Introduction}
Vitreous silica, the prototype material for the study of network forming
glasses, displays a number of anomalous behaviors both in the structural
and vibrational characteristics \cite{zar}. This is also true when this 
material is exposed to high pressures. Experiments have shown that 
at room temperature glass samples compressed beyond $\approx$ 12 GPa 
exhibit a permanent density increase of about 20 \% \cite{expe}. Both
experiments \cite{expe2} and numerical simulations \cite{simul,tse} have
shown that this irreversible densification is mainly due to an increase
of the average coordination number around the silicon atoms. An other
topic concerns the influence of the pressure on the vibrational spectrum
of vitreous silica. Very recently a substantial amount of experimental
work has been dedicated to the study of this influence and more precisely
to the effect of the densification on the so-called Boson peak (BP) 
\cite{sugai,arai,arai2}. The BP refers to an excess in the vibrational density
of states with respect to the Debye distribution and is located generally 
around 1.5 THz. It is not clear yet if the BP only shifts towards
higher frequencies upon densification \cite{sugai} or if this shift
is a result of the suppression of the BP \cite{arai}.
 With the use of numerical simulations using
a relatively realistic interaction potential one should be able to 
give a clear answer and connect this answer  to the structural transformations
observed in the afore mentioned studies \cite{expe2,simul,tse}. As far as
we know no simulations were done on the variation of the BP with the
density. Here we present such a study in a model silica glass using the
same pair-potential than Tse {\em et al.} \cite{tse}, the ``BKS'' potential 
\cite{bks}. This potential has also been used recently to study pressure 
induced amorphization of quartz \cite{badro} and seems appropriate 
to describe silica at high pressure. Our results showing a shift of the 
BP as the density increases between $2.2$ and $2.67$ g/$\rm{cm}^3$ are 
in relatively good agreement with the experimental Raman data and indicate 
a disappearance of the BP upon densification. In parallel we find similar
results as Tse {\em et al.} \cite{tse} concerning the structural changes 
around the silicon atoms which are therefore not shown again here.

\section{Simulations}

We perform classical (constant energy, constant volume) molecular-dynamics 
simulations on an assembly of 648 particles (216 Si$\rm{O}_2$ molecules) 
packed in a cubic box with periodic boundary conditions. The size of the box 
is fixed in order to study samples at the following densities: 2.17, 2.32, 
2.40 and 2.67 g/$\rm{cm}^3$. The particles interact via
the BKS potential \cite{bks} with the same parameters as in our previous
studies \cite{juju}. The low temperature glassy samples at a given density 
are obtained after a quench from the liquid state ($T \approx 7000$K) at a 
constant quenching rate of $2.3 \times 10^{14}$K/s. It is worth noticing that
this procedure is different from the one used in \cite{tse} where a
compression at room temperature was used in order to obtain the high
density samples. After the quench, the samples at zero temperature are 
relaxed during 120000 timesteps (84 ps) (during which the temperature rises
only slightly) and finally the vibrational
spectrum $g(\nu)$ is obtained by direct diagonalization of the dynamical
 matrix. This direct diagonalization is the most computer time consuming since
one has to deal with a $1944 \times 1944$ matrix with no {\em a priori} 
symmetry. Moreover since the Coulomb interactions have not been cut off, 
this matrix can not be considered as sparse (to reduce the computer time it
should also be noted that we did not calculate the eigenvectors in this
simulation). The eigenvalues of the dynamical matrix can also be obtained
from the Fourier transform of the velocity-velocity autocorrelation function
but in the low frequency region of the spectrum (which is the one we are
mostly interested in) this technique is generally less reliable. 
For each density two independent liquid samples were used in
order to improve slightly the statistics of the  results. Because of the
finite size of the simulation box we have discrete  values of the frequency
$\nu$ which are located between $\approx$ 0.8 and 40 THz in agreement with a
previous study on the same system \cite{elliott}.  Nevertheless since we are
interested in the low frequency part of the  spectrum we will concentrate here
on the frequency region below 10THz. 

\section{Results}

A standard way of extracting the BP from the vibrational spectrum 
is to plot $g(\nu)/\nu^2$ since in the Debye approximation $g(\nu) \approx 
\nu^2$ at low frequency. In fig.1 we show the averaged (over 2 samples)
spectrum at different densities together with the neutron diffraction 
results \cite{arai2} when available (the curves have been arbitrarily
shifted for clarity). The solid lines in the figure are the fits of 
$g(\nu)/\nu^2$ by the following ``generalized Lorentzian'' functional form 
\cite{ldv}:
\begin{equation}
f(\nu)=f_0.\nu^n \cdot
 \frac{1}{[\nu^2+\nu_0^2]^m }   
   \label{eq:lorgen}
\end{equation}

where $f_0$, $\nu_0$, $m$ and $n$ are adjustable parameters (this functional
form is useful to describe the BP even though it has no direct physical
meaning). In this figure the excess of modes between 
1 and 2 THz is clearly visible. Moreover a broadening of the peak when
increasing the density can be seen in agreement with the experimental 
results. The increase of the peak position with increasing density is
less pronounced than in experiment because of finite size effects. 
Indeed due to the limited size of
our simulation box the number of frequencies ``available'' on the low 
frequency side  of the spectrum is small. This explains the 
overestimated peak position at 2.2 g/$\rm{cm}^3$ compared to the experimental
data as has been shown recently \cite{horbach}. We did the calculation
for one sample at 2.2 g/$\rm{cm}^3$ containing 1536 particles and obtained
very similar results except a small shift of the peak position towards 
lower frequencies. Increasing the number of particles by a factor 2 improves
barely the description of the peak, since the lower cut-off frequency is 
proportional to the inverse of the box size. In any case increasing the
system size leads to a {\em decrease} of the peak position and an {\em increase} of its intensity (see \cite{horbach}) while increasing the density leads
to an {\em increase} of the peak position and a {\em decrease} of its 
intensity as can be seen in fig. 1. The finite size effects are certainly 
responsible for the differences between the simulation and the experimental 
data but do not affect the general behavior with density. Since only 2
densities have been considered in the neutron diffraction study,  we decided 
to compare our results to the more numerous Raman data available 
\cite{sugai,arai}. When doing this one has to be careful in order to 
compare things that are comparable. The reduced Raman intensity $I_R$ is 
related to the vibrational density $g(\nu)$ via a frequency dependent 
function $C(\nu)$. Previous experimental studies \cite{sokolov} and a 
recent study combining experiments and MD simulations \cite{font} have 
indicated that $C(\nu) \approx \nu$ in the region of the BP which implies 
that $I_R \propto g(\nu)/\nu$. Hence in the following we use the 
function $g(\nu)/\nu$ and compare its characteristics to the available 
Raman data.  To put this comparison on more quantitative grounds we 
fit the curves $g(\nu)/\nu = f(\nu)$ by a functional form similar to the
one given in Eq. 1, and we extract the position of the maximum 
 $\nu_{max}$ as well as the intensity of the maximum $I_{max}$ from the fits.
 In fig.2a and 2b the values of $\nu_{max}$ and $I_{max}$ are plotted as a 
function of the density respectively and compared to experimental Raman 
values. In fig.2a, the 2 points reproduced from \cite{sugai} (open squares) 
which is a study of the vibrational properties of silica as a function of 
{\em pressure} were obtained assuming a 15 \% densification of the high 
pressure sample as stated by the authors.

\section{Discussion}

The results reported in fig.2 show clearly that the BP is shifted towards
 higher frequencies and its intensity vanishes when the density of our model 
silica glass increases. In fact, compared to fig.1, not surprisingly, 
the relative variations of the peak position and intensity are more 
important when $g(\nu)/\nu$ is used. In fig.2a we see that the overall 
trend of the BP position as a function of density is correct even though 
our MD simulations overestimate this position compared to the experimental 
data which is probably due to finite size effects as stated earlier. 
Nevertheless it appears that our results come closer to the experimental 
data at high density even though the positive curvature seems to be absent 
in the simulation (the error bars in fig.2a and 2b result from the dispersion 
over the 2 samples studied here 
and therefore should be considered as a rough estimate of the error). This
shows that the shift of the BP towards higher frequencies is more important
than the artificial drift due to the fact that with increasing density
(decreasing system size) the lower cut-off frequency increases in our 
simulations. This increase would imply a worse description of the BP region 
but since the BP shifts towards higher frequencies our description becomes 
(relatively) better at high density. Another point concerns the influence
of the rapid quenching rate: indeed Vollmayr {\em et al.} have shown that
the height and to a lesser degree the shape of the BP depend on the
cooling rate \cite{vollmayr}. Nevertheless, similarly to the finite size 
effects, these authors show that a smaller cooling rate would lead to a 
behavior of the BP opposite to the one observed with increasing density. 
Therefore since all the samples have been quenched at the same rate the
evolution of the BP with the density can not be linked to cooling rate
effects. In fact all the previous studies show that using larger system 
sizes and smaller cooling rates would lead to a better agreement of the 
curves plotted in fig.2a especially at low density but would not affect
the behavior of the BP with the density which is the aim of this study.\\
Differences can also be seen between the experimental studies.  The principal 
distinction between the two experimental procedures is the temperature 
since Sugai {\em et al.}  performed the densification at room temperature 
while Inamura {\em et al.} compressed the samples at 700 C.
This might explain why our low temperature simulations are closer to the
experimental results obtained at room temperature. In order to check
the effect of the temperature, we performed some calculations on samples
quenched at 300K but no major differences were noticeable. Also we tried
a ``cold compression'' at 300K followed by a long relaxation run (200 ps)
but again the spectra were similar to the ones reported in fig.1.
The results in fig.2b confirm that with increasing density the BP 
tends to disappear. Again we do not find the almost linear decrease observed
in experiment at low density but such a behavior can be observed for the
higher densities (this shows again that our description becomes
better at high densities). Since we fixed the calculated and experimental
intensity of the BP to 1 at the usual density (2.2 g/$\rm{cm}^3$) for which 
we know that the calculated intensity is underestimated because of the 
finite size of the simulation box, we have a systematic overestimation
of the relative intensity of the BP at higher density. Nevertheless the 
overall trend is clear and it is coherent with the 
data of Inamura {\em et al.} \cite{arai2} who recently showed not only a 
shift towards high frequencies but also a disappearance of the BP. In their 
publication these authors attributed the suppression of the 
low-energy dynamics to the shrinkage of the 
void space by densification \cite{elliott2}. This analysis is coherent with
the soft potential model (the soft modes in the void space have been
suppressed by shrinkage) but seems to contradict the analysis of
the shift of the BP being due to phonon scattering by local density
fluctuations which would result in a shift of the BP only \cite{sugai}. 
Nevertheless the microscopic model of the BP is not settled yet
since the structural entities involved in the soft potential model have
not been clearly identified. The connection between this
study and the previous numerical studies investigating the structural 
changes under densification should permit to progress in such an 
identification.
 
\section{Conclusion}

We performed classical molecular-dynamics simulations combined with
the diagonalization of the dynamical matrix to investigate the influence
of the density on the Boson Peak in a model silica glass. Even though we
used a relatively small system size and a rather fast cooling rate, 
the results compare relatively well with experimental Raman data and show 
that upon densification the Boson Peak shifts to higher frequencies while
its intensity decreases. These results confirm the disappearance of the
BP in the high density samples as shown in a recent experimental 
study \cite{arai2} which could be due to a shrinkage of the void
space \cite{elliott2}. 
The quantitative agreement between our simulations and experiment is not 
perfect (it should be noted that differences exist also between the
experimental data) but we know the reasons of this discrepancy.
Nevertheless the BKS potential is able to give the correct behavior of the 
vibrational spectrum under compression and this shows once more the good 
quality of this pairwise interaction. Of course since we did not calculate the 
eigenvectors we have no informations on the nature of the modes dissapearing
under compression. But the disappearance of the excess of low-energy modes
under densification  is the first step in the clear identification of what the
vibrating  structural entities are. To perform this identification in a
simulation one needs tools more accurate than the radial pair distribution
function or the Vorono\"\i\ tessellation. Also if the connection between the
BP and the plateau in the thermal conductivity exists  one
should see a variation of the plateau with the density: this variation has
already been reported experimentally \cite{zhu} but not yet confirmed in
a numerical simulation.

\section{Acknowledgments}
We would like to thank M. Foret and C. Levelut for helpful discussions
on this work. Part of the simulations have been done at the Centre 
Informatique  National de l' Enseignement Superieur (Montpellier).
  


\begin{figure}
\psfig{file=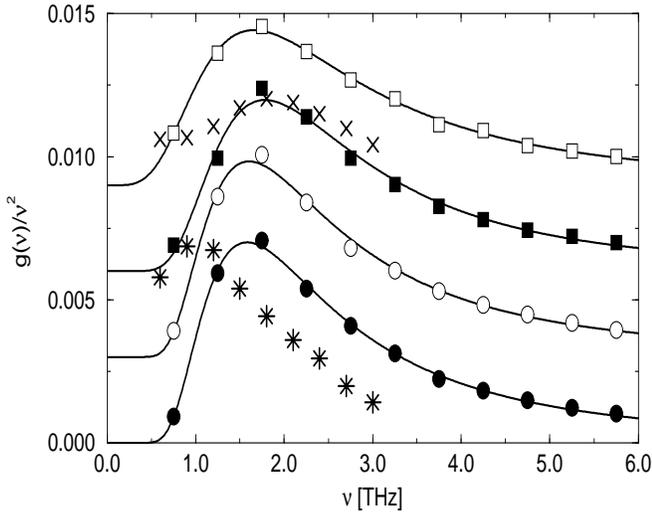,width=8.6cm,height=7.cm}
\caption{Plot of $g(\nu)/\nu^2$ versus $\nu$ as a function of density:
$\bullet$: 2.17 g/$\rm{cm}^3$; $\circ$: 2.32 g/$\rm{cm}^3$; 
\protect\rule[0.5pt]{2.0mm}{2.0mm}: 2.40 g/$\rm{cm}^3$; $\Box$: 2.67 g/$\rm{cm}^3$. Comparison with experimental neutron diffraction data \cite{arai2} (the intensity at 2.2 g/$\rm{cm}^3$ has been adjusted to the corresponding MD 
intensity ): $\ast$: 2.2 g/$\rm{cm}^3$; 
$\times$: 2.63 g/$\rm{cm}^3$.\\
The solid lines correspond to the fit of the MD results with the ``generalized 
Lorentzian'' given in Eq. 1 (see text)
}
\label{Fig.1}
\end{figure}

\vspace*{-0.6cm}
\begin{figure}
\psfig{file=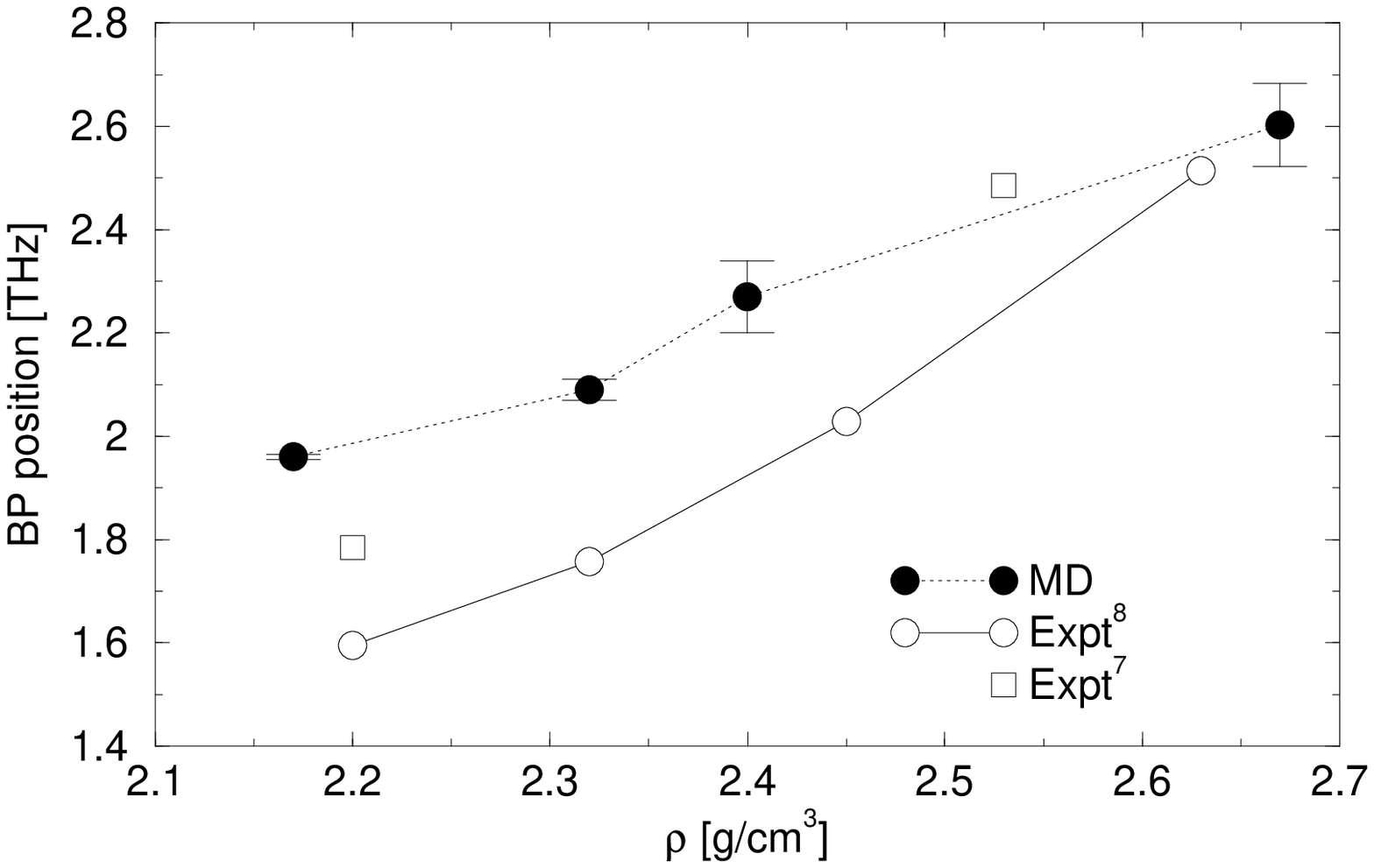,width=8.5cm,height=7.cm}
\psfig{file=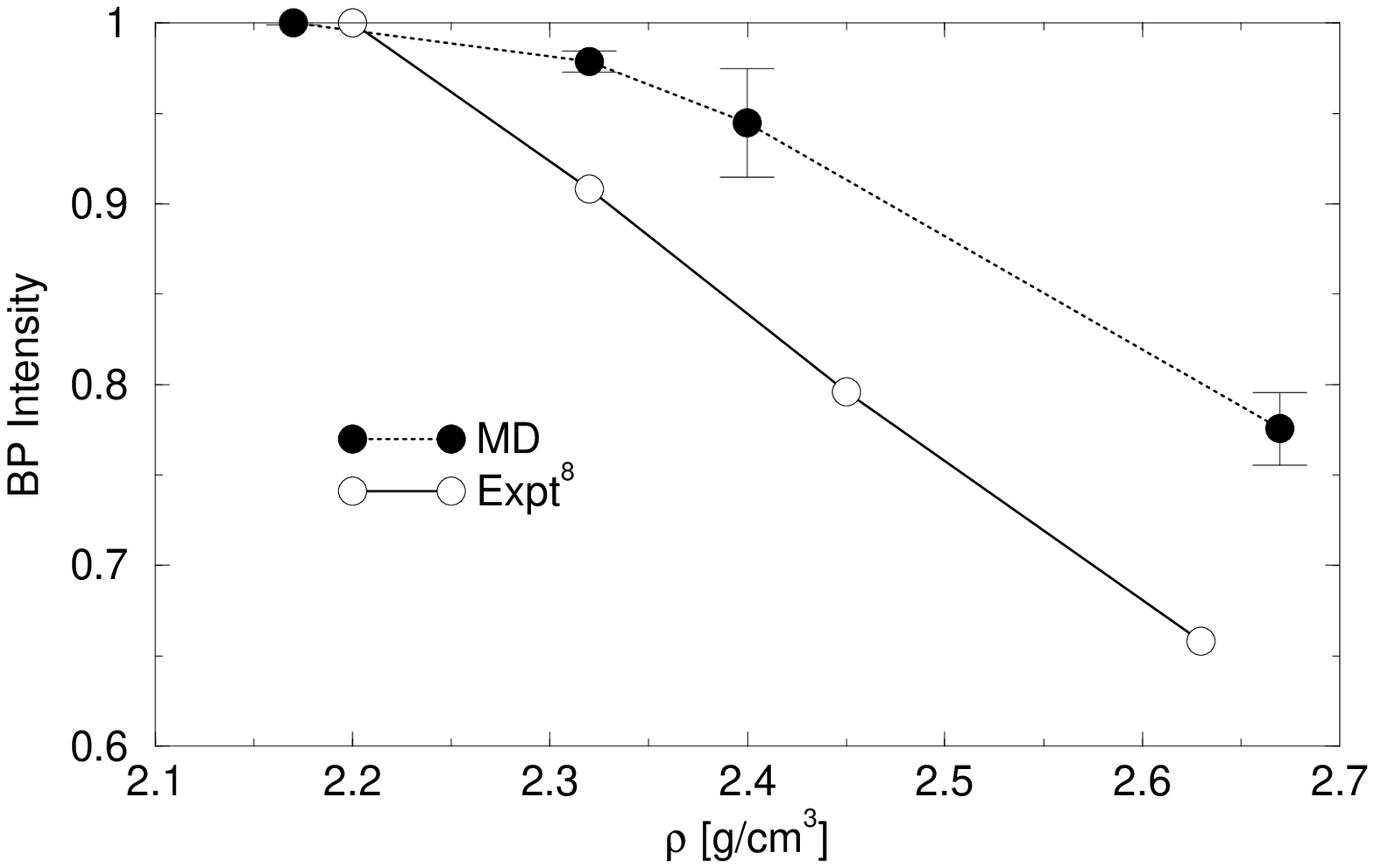,width=8.5cm,height=7.cm}
\caption{(a) Variation of the position of the BP as a function of density;
(b) Variation of the intensity of  the BP as a function of density.}
\label{Fig. 2a}
\end{figure}

\end{document}